\documentclass{article}
\usepackage[utf8]{inputenc} 
\usepackage[T1]{fontenc}    
\usepackage{hyperref}       
\usepackage{url}            
\usepackage{booktabs}       
\usepackage{amsfonts}       
\usepackage{nicefrac}       
\usepackage{microtype}      
\usepackage{lipsum}
\usepackage{fancyhdr}       
\usepackage{graphicx}       
\graphicspath{{media/}}     
\usepackage{xcolor}
\hypersetup{
    colorlinks,
    linkcolor={red!50!black},
    citecolor={blue!50!black},
    urlcolor={blue!80!black}
}
\usepackage{subfig}
\usepackage{algorithm}
\usepackage[noend]{algpseudocode}
\usepackage{algcompatible}
\usepackage{pgfplots}
\usetikzlibrary{shapes.geometric}
\pgfplotsset{compat=1.17}
\usepackage{tikz}
\usepackage{wrapfig}
\usepackage{pifont}
\usepackage{tablefootnote}
\usepackage{amsmath}
\newcommand{\cmark}{\ding{51}}%
\newcommand{\xmark}{\ding{55}}%
\newcommand{\cmmnt}[1]{}
\usetikzlibrary{positioning}

\usepackage{PRIMEarxiv}

\title{End-to-end audio strikes back: Boosting Augmentations towards an Efficient Audio Classification Network}

\author{
  Avi Gazneli, Gadi Zimerman, Tal Ridnik, Gilad Sharir, Asaf Noy \\
  DAMO Academy, Alibaba Group \\
  \texttt{\{avi.g, gadi.zimerman, tal.ridnik, gilad.sharir, asaf.noy\}@alibaba-inc.com} \\
}

\begin{document}
\maketitle

\begin{abstract}
While efficient architectures and a plethora of augmentations for end-to-end image classification tasks have been suggested and heavily investigated, state-of-the-art techniques for audio classifications still rely on numerous representations of the audio signal together with large architectures, fine-tuned from large datasets. By utilizing the inherited lightweight nature of audio and novel audio augmentations, we were able to present an efficient end-to-end (e2e) network with strong generalization ability. Experiments on a variety of sound classification sets demonstrate the effectiveness and robustness of our approach, by achieving state-of-the-art results in various settings. Public code is available at: \href{https://github.com/Alibaba-MIIL/AudioClassfication}{https://github.com/Alibaba-MIIL/AudioClassification}.
\end{abstract}


\section{Introduction}
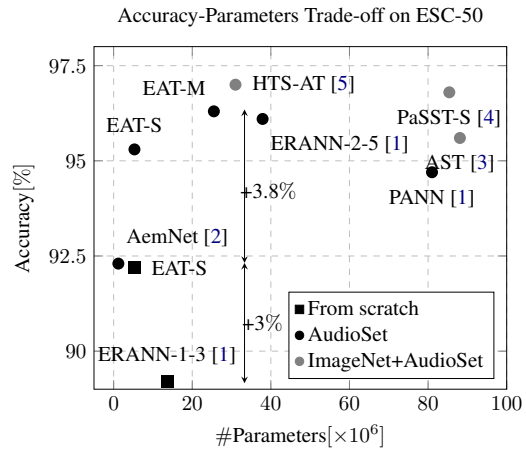
\begin{wrapfigure}{r}{0.45\textwidth}
    \centering
        \resizebox{200pt}{!}{%
            \begin{tikzpicture}
            
            \draw [stealth-stealth](2.5,0.1) -- (2.5,2.1);
            \draw [stealth-stealth](2.5,2.1) -- (2.5,4.65);
            \node at (2.8,1.1) {+3$\%$};
            \node at (2.9,3.3) {+3.8$\%$};
            \begin{axis}
                    [
            title={Accuracy-Parameters Trade-off on ESC-50},
            xlabel={$\#$Parameters$[\times10^6]$ },
            ylabel={Accuracy$[\%]$ },
            xmin=-5, xmax=100,
            ymin=89, ymax=98,
            xtick={0,20,40,60,80,100,120},
            ytick={85, 87.5, 90, 92.5, 95, 97.5, 100},
            legend pos=south east,
            xmajorgrids=true,
            ymajorgrids=true,
            grid style=dashed,
            legend cell align={left},]


        \addplot[ only marks,mark size=2pt, mark=square*, mark options=black] coordinates {(0,0)};
 
            \node[label={[label distance=0.05cm]90:{ERANN-1-3 \cite{kong2020panns}}},fill,inner sep=3pt] at (axis cs:13.6,89.2) {};
            \node[label={[label distance=0.05cm]0:{EAT-S}},fill,inner sep=3pt, color=black] at (axis cs:5.3,92.2) {};

            \addlegendentry{From scratch}
        
        \addplot[ only marks, , mark options=black] coordinates {(0,0)};
            \node[label={[label distance=0.05cm]270:{PANN \cite{kong2020panns}}},circle,fill,inner sep=2pt] at (axis cs:81,94.7) {}; 
            \node[label={[label distance=0.05cm]275:{ERANN-2-5 \cite{kong2020panns}}},circle,fill,inner sep=2pt] at (axis cs:37.9,96.1) {}; 
            \node[label={[label distance=0.05cm]90:{EAT-S}},circle,fill,inner sep=2pt, color=black] at (axis cs:5.3,95.3) {};
            \node[label={[label distance=0.05cm]95:{EAT-M}},circle,fill,inner sep=2pt, color=black] at (axis cs:25.5,96.3) {};
            \node[label={[label distance=0.05cm]85:{AemNet \cite{lopez2021efficient}}},circle,,fill,inner sep=2pt] at (axis cs:1.2,92.3) {};
            \addlegendentry{AudioSet}

        \addplot[ only marks,mark options=gray] coordinates {(0,0)};
            \node[label={[label distance=0.05cm]270:{AST \cite{gong2021ast}}},circle,fill,inner sep=2pt, color=gray] at (axis cs:88.1,95.6) {};
            \node[label={[label distance=0.05cm]270:{PaSST-S \cite{koutini2021efficient}}},circle,fill,inner sep=2pt, color=gray] at (axis cs:85.4,96.8) {};  
            \node[label={[label distance=0.05cm]0:{HTS-AT \cite{chen2022hts}}},circle,fill,inner sep=2pt, color=gray] at (axis cs:31,97) {};  
            \addlegendentry{ImageNet+AudioSet}
            


            \end{axis}

            \end{tikzpicture}
            }
            \captionsetup{justification=centering}

            \caption{Comparison of our EAT architecture. Achieving SotA on 'From scratch' training (+3$\%$ from \cite{verbitskiy2021eranns}) and on AudioSet pretrain (+3.8$\%$ from efficient e2e method \cite{lopez2021efficient}) }

            \label{tikzpicture:graph_esc50_pretrained}
\end{wrapfigure}

\begin{figure}[htb]
    \centering
    \subfloat[]{\includegraphics[width=.45\textwidth]{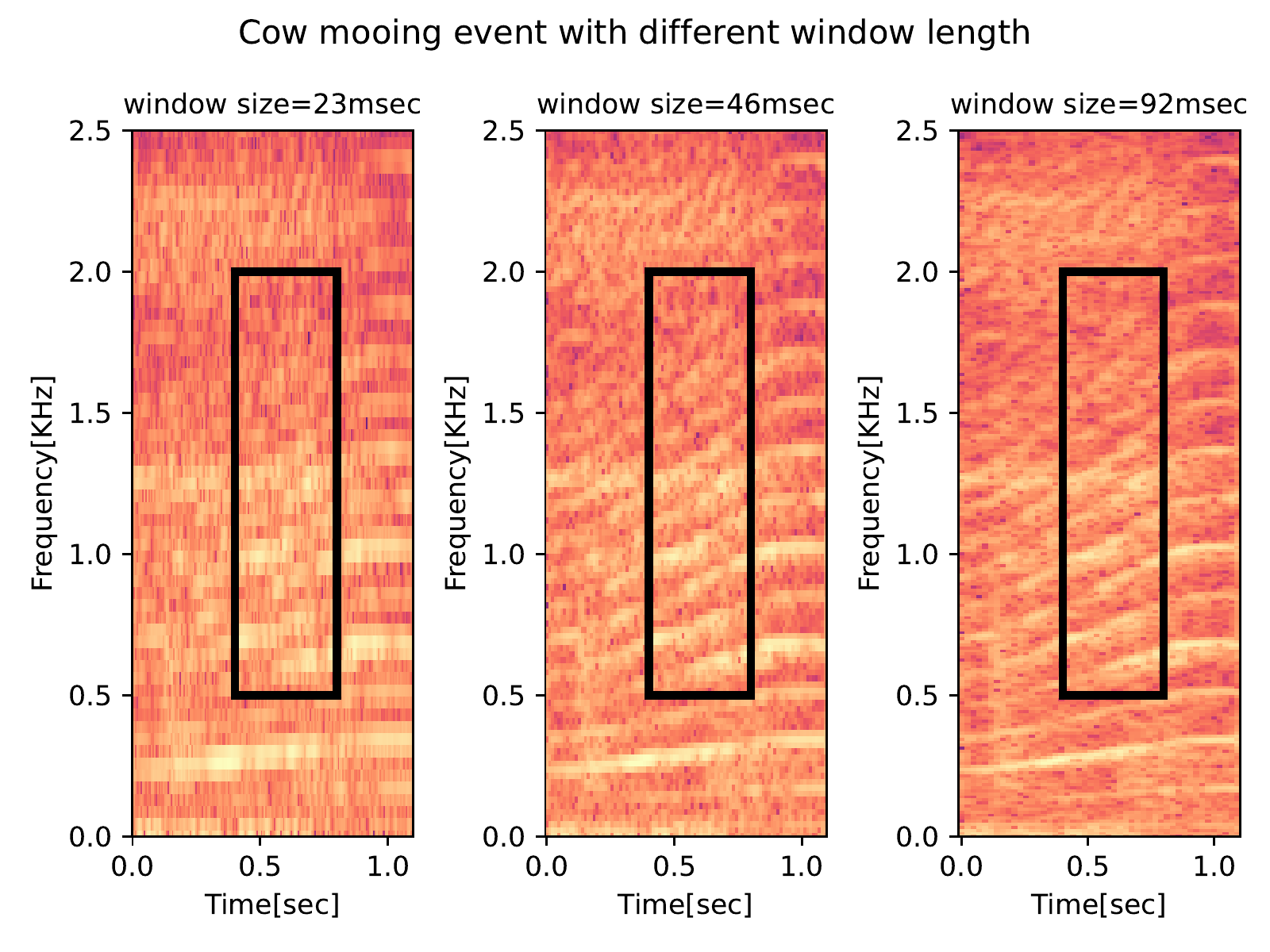}}
    \qquad
    \subfloat[]{\includegraphics[width=.45\textwidth]{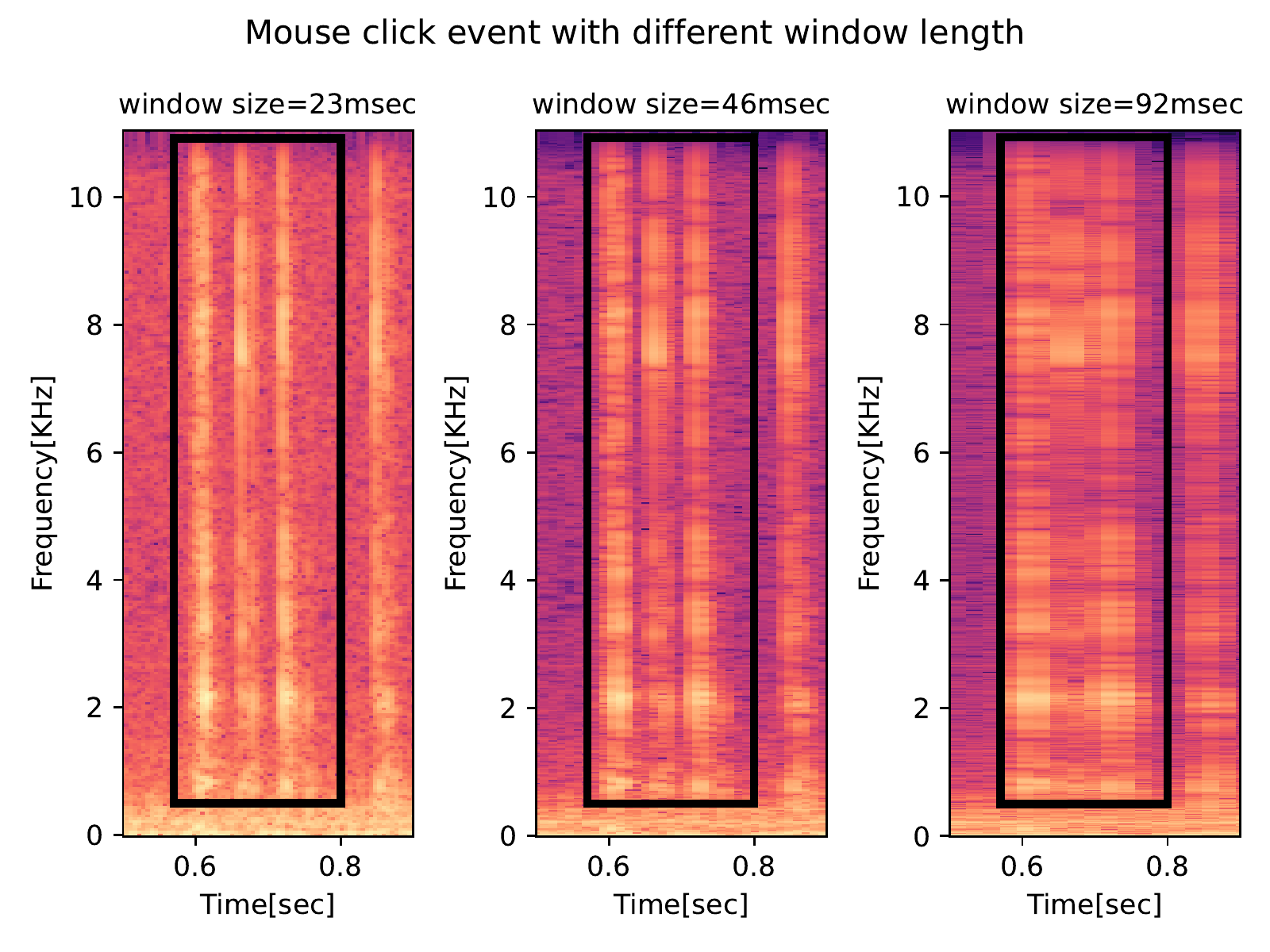}}
    \caption{Impact of window length on time-frequency resolution: Observing a slow event, cow mooing (a), compared to a fast event, mouse click (b) through 3 typical window lengths, with $75\%$ overlap.}%
    \label{fig:mouse-cow-mres}%
\end{figure}

In signal processing, sound pattern recognition plays a crucial role with a wide range of applications. Recognition can be modeled as a classification task, whether \textit{single-label} or \textit{multi-label}, where the algorithm outputs predictions for class labels. A typical audio signal consists of speech, music, and other environmental sounds. Environment sound refers to a wide range of classes spanning from sea waves through engines, etc. An audio signal is typically handled by converting it to a known time-frequency (T-F) representation, usually with the help of the \textit{spectrogram} and its compressed form, known as the \textit{mel-spectrogram}. The former is obtained by applying \textit{Short-Time Fourier Transform} (STFT) on the waveform and taking the magnitude, while the latter requires an additional stage in which mel filter-banks are applied for squashing the frequency axis to mel bins with logarithmic spacing. However, the use of mel-spectrogram comes at the cost of having to carefully adjust the parameters for time-frequency resolution and compression rate, which might vary for different classes. By nature, sound samples for distinct classes possess different characteristics manifested mainly in duration and frequency spectrum. Hence, finding one set of parameters suitable for all is implausible. For instance, the duration of the mouse-click event lasts several milliseconds, necessitating a shorter window size, compared to the cow mooing event, which lasts a few seconds, as depicted in Fig. \ref{fig:mouse-cow-mres}.

\begin{figure}[htb]
    \centering
    \includegraphics[width=.45\textwidth]{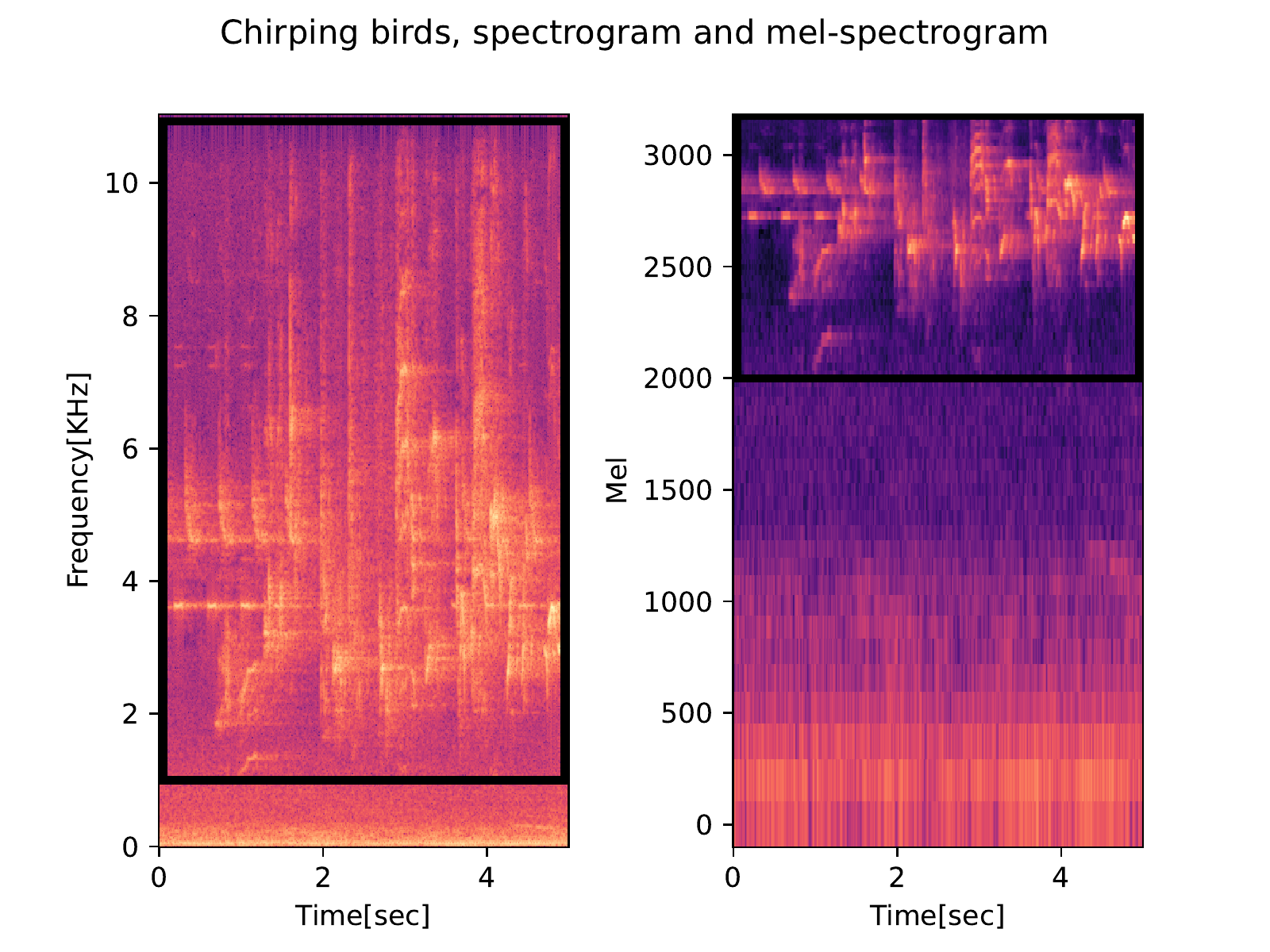}
    \caption{Impact of mel scale compression: Comparing linear frequency spacing (left) against logarithmic mel scale (right).}
    \label{fig:mel-linear2}
\end{figure}

\noindent In addition, the usage of frequency compression with logarithmic binning can also degrade the signal. For instance, chirping bird sounds naturally occupy a high-frequency band, where the transformation assigns coarsely spaced bins, as depicted in Fig. \ref{fig:mel-linear2}, which was already handled by \cite{guzhov2021esresnet, verbitskiy2021eranns}.

In audio processing, the representation issue remains an active topic. The trade-off travels from domain knowledge incorporation in signal representations at the expense of complex architectures requiring large data to fit, through end-to-end systems which maintain performance gap, mainly on limited data scenarios \cite{pons2017end, lopez2021efficient, tokozume2017learning}.\\
In this work we prefer the end-to-end strategy. The absence of pre-processing streamlines the system, by requiring fewer parameters to tune, and facilitates shifting between tasks with distinct signal content. Furthermore, usage of raw signal allows us to apply wide set of augmentations, including two novel schemes, which significantly decrease the gap reported in the early works. On top of data manipulation schemes, we propose a neural network, designed to handle raw audio signal characteristics. The resulted solution is simple with a low memory footprint and short inference time, as well as robust for distinct audio contents. The proposed system was evaluated on several public datasets, such as ESC-50 \cite{piczak2015esc}, UrbanSound8K \cite{salamon2014dataset}, AudioSet \cite{gemmeke2017audio} and SpeechCommands \cite{warden2018speech}, achieving state of art results on several scenarios.\\
The contribution of the paper can be summarized as follows:
\begin{itemize}
    \item Introducing two novel and effective augmentations for audio signals
    \item Designing an efficient deep learning architecture
    \item Demonstrating the potential of end-to-end methods, and their superiority in several audio benchmarks
\end{itemize}

\section{Related Work}
Audio pattern recognition systems mainly rely on transforming the raw-audio signal to time-frequency representation, mostly to a mel-spectrogram representation, and then using deep neural networks to output the class prediction. Empirically, they outperform early models which used e2e audio networks \cite{lee2017sample, tokozume2017learning, guzhov2021esresnet}. \cite{huzaifah2017comparison} performed a comprehensive comparison among numerous representations given the network architecture, deducing the superiority of the mel-spectrogram compared to others. The common modus operandi in deep learning is to use transfer learning from pretrained networks. \cite{palanisamy2020rethinking} demonstrated the benefit of using architectures such as DenseNet \cite{iandola2014densenet}, ResNet \cite{he2016deep} and Inception \cite{szegedy2015going} pre-trained on ImageNet \cite{deng2009imagenet} when applied on mel-spectrograms for the audio classification tasks. To adapt the input type to mentioned architectures, they deduced that incorporating different time-frequency resolution maps is beneficial over simple replication across the channels. The authors of \cite{guzhov2021esresnet, guzhov2021esresnextfbsp} preferred the log-spectrogram and wavelet-based representations for their ResNet50-based network. The diversity in input type may indicate on lack of robustness across tasks and require carefully adjusting the pre-defined parameters.
\newline Several works focused on e2e architectures, albeit, there was a performance gap mainly on limited data scenarios \cite{pons2017end, tokozume2017learning, lopez2021efficient}. Improved performance obtained by incorporating some domain knowledge with initialization \cite{tax2017utilizing, guzhov2021esresnextfbsp}, complex architectures \cite{kong2020panns}, even using additional self supervised training phase \cite{wang2021multi}.\\
Lately, the emergence of transformers \cite{parmar2018image} infiltrated to audio processing domain with an invigorating effect. For instance, \cite{gong2021ast} applied the transformer on mel-spectrogram patches with impressive results across several datasets. In order to relax the training complexity and enabling variable size inference, \cite{koutini2021efficient} suggested dedicated regularization scheme during the training process together with the disentanglement of positional encoding to time and frequency axis. However, the state-of-the-art (SotA) results rely on complex models, posing hard constraints on deployment and inference.\\
\cite{park2019specaugment} introduced a set of augmentations and became ubiquitous for spectrogram-based systems. The list involves time warping and masking the time/frequency axis. An additional widely used \cite{purwins2019deep, purwins2019deep, verbitskiy2021eranns, kong2020panns, gong2021ast, gong2021psla} strategy is to mix pairs of samples in amplitude \cite{tokozume2017learning}, similar to \cite{zhang2017mixup} on image pixels, except the mixing ratio normalized by sample gain. Our solution expands the augmentation portfolio by suggesting two novel mixing strategies, by scrambling the pairs of signals in frequency and phase, in addition to a neural network architecture dedicated to processing the signal efficiently.

\section{Method}
In this chapter, we describe our approach to audio classification. In general, the method involves augmenting data distribution and better integrating sound characteristics into architecture design. First, we will describe our architecture for audio classification, then we will present novel ways of augmenting sound signals. 

\subsection{EAT Architecture}
\begin{figure}[htb]
    \centering
    \captionsetup{justification=centering}
    \includegraphics[width=.35\textwidth]{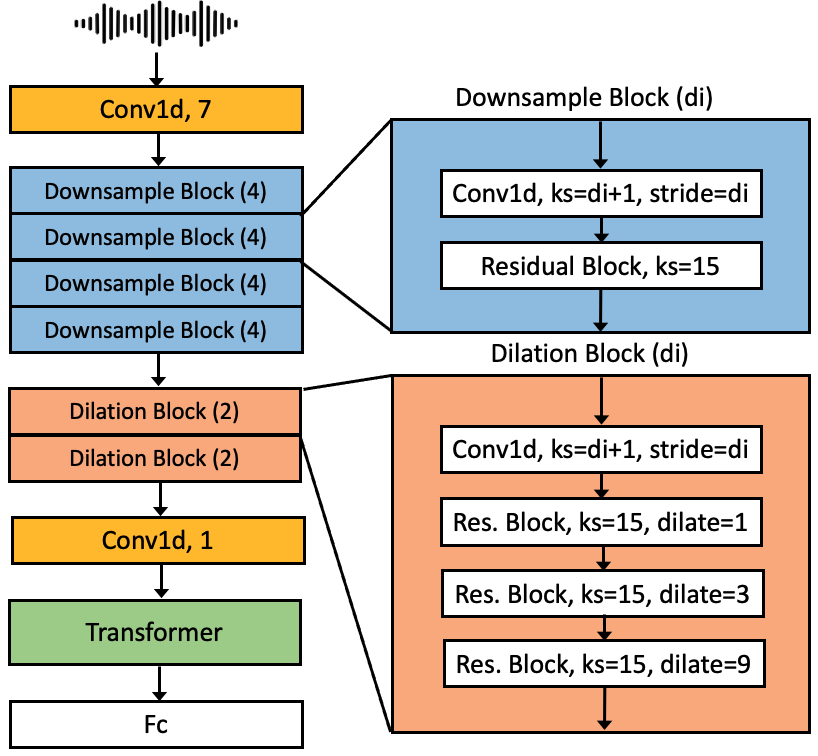}
    \caption{The proposed EAT architecture, CNN-\\ style backbone followed by a transformer.}
    \label{fig:arch}
\end{figure}

The proposed audio classification network is shown in Fig. \ref{fig:arch}. During this stage, the primary focus was to build a neural network that has a large receptive field, while keeping complexity low. One can decompose the network into two main blocks, a 1D convolution stack, and a transformer encoder block. The former downsamples along the time axis with a convolution layer coupled to a fixed low-pass filter \cite{zhang2019making, ridnik2021tresnet}, followed by intermittent residual blocks \cite{he2016deep}. The residual blocks are modified according to \cite{you2021axial}, consisting of depth-wise convolution with a large kernel operating on the time axis, and $f(x)$ is convolution with kernel size equal to 1 operating across channels. At this point, the signal is decimated using a sequence of factors ${d_i}$ by an overall factor of $d=\prod d_i$. For instance, signal with a 5-second duration the downsampling sequence is equal to $[4,4,4,4]$, performing a reduction by a factor of $256$. This can be to some extent linked to downsampling performed during spectrogram operation\footnote{Depending on the sampling rate and STFT parameters, for example, the typical choice for 22.05KHz can be window size equal to 1024 with hopping of 256, which effectively decimates by 256 the time axis.}. The following building blocks perform additional reduction, with each followed by a stack of dilated residual blocks \cite{kumar2019melgan}. This refinement enables to increase in the receptive field per frame, hence being more robust to variable duration events among the classes in environmental sound scenarios. Gathering feature maps across frames was implemented using a transformer encoder block, which followed by fully connected layer to project the embedding vector to class space. For complexity analysis and details about EAT-S and EAT-M models refer to Appendix \ref{app:arch}.

\subsection{Data Augmentation}
Data augmentation is a ubiquitous step during the training phase for deep learning networks, particularly in the case of limited data. \cite{pons2017end, tokozume2017learning, lopez2021efficient} already pointed out the inferiority of e2e audio-based systems in case of limited data. This can be mitigated to some extent by enriching the list of augmentations. We noticed that specifically, augmentations involving label mixing were beneficial for generalization. We suggest mixing frequency bins and phases and naming them as \textit{FreqMix} and \textit{PhaseMix}. In FreqMix, alg. $[$\ref{alg:freqmix}$]$, given pair of samples with corresponding labels, we perform the mix by choosing low-frequency bins from one sample and concatenating with high frequency with another sample, or vice versa. The operation is equivalent to applying ideal filters in frequency domain and adding their results. In addition to the mixup \cite{zhang2017mixup} operation, FreqMix contributes to enlarging the size of in-between samples, by introducing a convex combination of filtered versions of the pair, Equation \ref{eq:freqmix}. The filtering erases a small amount of information in the frequency domain from the original sample, and fills the filtered spectrum portion from its counterpart. Sound signals are rarely narrow-band, such as a pure sine wave, thus removing a segment of contiguous frequency is unlikely to erase the entire data. The mixing, in addition spans the linear behaviour for in-between samples, to larger set than the original mixup.
\begin{align}\label{eq:freqmix}
    \text{\textbf{MixUp}: }& x_{mix}[n]=\lambda \cdot x_1[n] + (1-\lambda) \cdot x_2[n]\nonumber\\
    \text{\textbf{FreqMix}: }& x_{mix}[n]=x_1[n]*h_1[n;\lambda] + x_2[n]*h_2[n;\lambda]
\end{align}
where, $h_1[n;\lambda]$, $h_2[n;\lambda]$ are low-pass and high-pass filters parametrized by $\lambda$, controlling the cut-of frequency.\\
Raw audio maintains an additional signal characteristic, the phase. To demonstrate that phase contains some amount of discriminative information, we conducted an experiment, as described in Appendix \ref{app:phase}, in which we synthesized waveforms from the phase of the signal and trained our neural network. The classification result was significantly higher than the random guess, as detailed in Table \ref{tab:phonly}. As a consequence, we suggest adding more robustness by mixing the phase among the samples, and name it PhaseMix. In PhaseMix, alg. $[$\ref{alg:phmix}$]$, the amplitude of the original signal remains, while phase is mixed. The level of mixing is dictated according to the mixing ratio, which is randomly drawn in each iteration.\\
These two mixing strategies come on top of applying modified mixup \cite{tokozume2017learning} and cutmix \cite{yun2019cutmix} that we adopted for the 1D case, with results summarized in Table \ref{tab:ablations}.\\
In addition to the transformations involving the mixing of labels, we used transforms that preserved labels, such as amplitude manipulation, re-sampling, filtering, time-shifting, and a variety of noises. It is worth mentioning that working with raw signals enables incorporating larger set of transformations with mathematical interpretations. Shifting the signal in time, for example, is reflected in the frequency domain by adding linear phase. This has no affect when working with spectrogram, where the magnitude step by definition discards phase. Furthermore, mimicking the time shift by shifting the spectrogram along the time axis, is not an equivalent operation, since time shift and absolute value are not interchangeable.\\
With the proposed augmentation scheme, the early reported gap for limited data scenarios \cite{pons2017end, tokozume2017learning, lopez2021efficient} was eliminated. The whole set of transforms detailed in Table \ref{tab:label-augs}.

\begin{algorithm}[htb]
\caption{FreqMix}
\label{alg:freqmix}
\begin{algorithmic}[1]
    \STATE let $(x_1, y_1)$, $(x_2, y_2)$ be samples from dataset X
    \STATE $X_1=STFT(x_1)$,
    \STATE $X_2=STFT(x_2)$ 
    \STATE $\lambda \sim U[0.5, 1]$, \quad $p \sim U[0,1]$
    \STATE $k_c=int(\lambda \cdot n_{fft})$
    \STATE $X_{mix} = \begin{cases}
                        X_1[n_{fft}-k_c:,:] \oplus X_2[:k_c,:] &
                        p\leq 0.5\\
                        X_1[:k_c,:] \oplus X_2[n_{fft}-k_c:,:] & p > 0.5
                      \end{cases}$
    \STATE $x_{mix}=ISTFT(X_{mix})$
    \STATE $y_{mix}=\lambda \cdot y_1 + (1-\lambda)\cdot y_2$
\end{algorithmic}
\end{algorithm}

\begin{algorithm}[htb]
\caption{PhaseMix}
\label{alg:phmix}
\begin{algorithmic}[1]
    \STATE let $(x_1, y_1)$, $(x_2, y_2)$ be samples from dataset X
    \STATE $X_1=STFT(x_1)$
    \STATE $X_2=STFT(x_2)$
    \STATE $\phi_1[k,l]=\angle X_1[k,l]$
    \STATE $\phi_2[k,l]=\angle X_2[k,l]$
    \STATE $\lambda \sim U[0, 1]$ 
    \STATE $\lambda_y=0.5\cdot \lambda + 0.5$
    \STATE $\phi_{mix}=\lambda \cdot \phi_1 + (1-\lambda) \cdot \phi_2$
    \STATE $X_{mix}[k,l]=\left|X_1[k,l]\right|\cdot
            e^{j\phi_{mix}[k, l]}$
    \STATE $x_{mix}=ISTFT(X_{mix})$
    \STATE $y_{mix}=\lambda_y \cdot y_1 + (1-\lambda_y) \cdot y_2$
\end{algorithmic}
\end{algorithm}

\begin{table}[htb]
\centering
\caption{List of label preserving and label mixing augmentations}
\label{tab:label-augs}
\begin{tabular}{cc}
    \toprule
    Name & Description\\
    \midrule
    amplitude & random amplitude to whole or fragment of sample\\ 
    noise     & white, blue, pink, violet, red, uniform, phase noise\\
    time shift & linear/cyclic with integer and  fractional delay\\
    filtering & low/high pass filter with randomized  cutoff frequency\\
    invert polarity & multiply by -1\\
    time masking & similar to cutout \cite{devries2017improved}, masking fragment of sample\\
    quantization & quantize sample using $\mu$ law or linear regime\\
    \midrule
    mixup & mixing amplitude \cite{zhang2017mixup}, \cite{tokozume2017learning}\\
    timemix & mixing in time axis, similar to cutmix \cite{yun2019cutmix}\\
    freqmix & alg.$[$\ref{alg:freqmix}$]$\\
    phasemix & alg.$[$\ref{alg:phmix}$]$\\
    \bottomrule
\end{tabular}
\end{table}

\section{Experiments}
In this section we will provide the experiments conducted on public classification benchmarks, such as ESC-50 \cite{piczak2015esc}, AudioSet \cite{gemmeke2017audio}  UrbanSound8K \cite{salamon2014dataset}. In addition to ESC scenarios, we examined the system on SpeechCommands \cite{warden2018speech} dataset, containing spoken words in English, to show some robustness to audio signal type. During the training process, we used AdamW \cite{loshchilov2017decoupled} optimizer with maximal learning rate of $5\cdot10^{-4}$ and one cycle strategy \cite{smith2018disciplined}. In addition, we use weight decay with $10^{-5}$, EMA \cite{tarvainen2017mean, izmailov2018averaging} with decay rate of 0.995 and SKD \cite{zhang2019your}. The loss is label-smoothing with a noise parameter set to $0.1$ for single-label classification tasks, and binary cross-entropy for the multi-label classification case. When applying mixing augmentations we use multi-label objective and use binary cross-entropy, as suggested by \cite{wightman2021resnet}. To handle distinct sample lengths across datasets we adjust the parameter controlling the downsample of the network. The set of augmentations used is detailed \ref{tab:label-augs}, with one noise type and one mixing strategy being randomly selected in each iteration.

\subsection{ESC-50}
The \textit{ESC-50} set \cite{piczak2015esc} consists of 2000 samples of environmental sounds for 50 classes. Each sample has a length of $5$ seconds and is sampled at 44.1KHz. The set has an official split into $5$ folds. We resampled the samples to 22.05KHz for being compliant with the majority of other works, and followed the standard $5$-fold cross-validation to evaluate our model. Each experiment repeated three times and averaged to final score.

\begin{table}[H]
\centering
  \caption{ESC-50, accuracy with model size and inference time measured on P-100 machine.}
  \label{tab:esc50-scrach}
  \begin{tabular}{ccccccl}
    \toprule
    Model & e2e & Pretrained & Accuracy$[\%]$ & $\#$Parameters$[\times10^6]$ & time$[msec]$\\
    \midrule
    ESResNet-Att \cite{guzhov2021esresnet} & \xmark & none & 83.15 & 32.6 & 11.3\\
    ERANN-1-3 \cite{verbitskiy2021eranns} & \xmark & none & 89.2 & 13.6 & -\\
    EnvNet-v2 \cite{tokozume2017learning} & \cmark & none & 84.9 & 101 & 2.7\\
    AemNet WM$1.0$ \cite{lopez2021efficient} & \cmark & none & 81.5 & \textbf{5} & -\\
    EAT-S & \cmark & none & \textbf{92.15} & \textbf{5.3} & 8.3\\
    \midrule
    PANN \cite{kong2020panns} & \xmark & AudioSet & 94.7 & 81 & - \\
    ERANN-2-5 \cite{verbitskiy2021eranns} & \xmark & AudioSet & \textbf{96.1} & 37.9 & -\\
    AemNet-DW WM$1.0$ \cite{lopez2021efficient} & \cmark & AudioSet & 92.32 & \textbf{1.2} & -\\
    EAT-S & \cmark & AudioSet & 95.25 & 5.3 & 8.3\\
    EAT-M & \cmark & AudioSet & \textbf{96.3} & 25.5 & 9.6\\
    \midrule
    AST \cite{gong2021ast} & \xmark & ImageNet+AudioSet & 95.6 & 88.1 & 26.7\\
    PaSST-S \cite{koutini2021efficient} & \xmark & ImageNet+AudioSet & \textbf{96.8} & 85.4 & 25.4\\
    HTS-AT \cmmnt{\tablefootnote{During inference the sample repeated twice rather zero padding to match the pretrained sequence length of 10 seconds}}\cite{chen2022hts} & \xmark & ImageNet+AudioSet & \textbf{97} & \textbf{31} & -\\
    \bottomrule
\end{tabular}
\end{table}

\noindent It is evident from the results that our method is more effective than others under the same settings. In absence of external data, the next in line in accuracy \cite{verbitskiy2021eranns} possess $\times 2.6$ more parameters\cmmnt{and presumably longer inference time}, while similar model size network \cite{lopez2021efficient} has a  $10\%$ gap in accuracy. In the AudioSet fine-tuned case, we manage achieve SotA while being $33\%$ lighter than \cite{verbitskiy2021eranns}.

\subsection{UrbanSound8K}
\textit{UrbanSound8K} is an audio dataset containing $8732$ labeled sound samples, belonging to $10$ class labels, split to $10$ folds. The samples last up to $4$ seconds and the sampling rate varies from $16$KHz-$48$KHz. The classes are drawn from the urban sound taxonomy and all excerpts are taken from field recordings \footnote{Can be found at \url{www.freesound.org}}. The experiment was conducted on the official 10 fold split, with samples resampled to 22.05KHz and zero-padding the short samples to 4 seconds. 

\begin{table}[htb!]
\centering
  \caption{UrbanSound8K, accuracy with model size and inference time measured on P-100 machine}
  \label{tab:urban8k}
  \begin{tabular}{ccccccl}
    \toprule
    Model & e2e & Pretrained & Accuracy$[\%]$ & $\#$Parameters$[\times10^6]$ & time$[msec]$\\
    \midrule
    ESResnet-Att \cite{guzhov2021esresnet} & \xmark & none & $82.76$ & $32.6$ & $11$\\
    AemNet WM$1.0$ \cite{lopez2021efficient} & \cmark & none & 81.5 & \textbf{5} & -\\
    ERANN-1-4 \cite{verbitskiy2021eranns} & \xmark & none & 83.5 & 24.1 & -\\
    EAT-S & \cmark & none & \textbf{85.5} & \textbf{5.3} & 8.5\\
    \midrule
    ERANN-2-6 \cite{verbitskiy2021eranns} & \xmark & AudioSet & \textbf{90.8} & 54.5 & -\\
    EAT-S & \cmark & AudioSet & 88.1 & \textbf{5.3} & 8.5\\
    EAT-M & \cmark & AudioSet & \textbf{90} & 25.5 & 9.6\\
    \midrule
    ESResNeXt-fbsp \cite{guzhov2021esresnextfbsp} & \xmark & ImageNet+AudioSet & 89.14 & 25 & $18.5$\\
  \bottomrule
\end{tabular}
\end{table}

\noindent The results on the UrbanSound8K dataset, detailed in Table \ref{tab:urban8k}, follow the same pattern as on the ESC-50, Table \ref{tab:esc50-scrach}, by outperforming previous approaches in the limited data scenario, while being competitive in the fine-tune mode.

\subsection{SpeechCommands}
\textit {Speech Commands V2} \cite{warden2018speech} is a dataset consisting of $\sim106K$ recordings for 35 words with a 1-second duration, with a sampling rate equal to 16KHz. The set has an official train, validation, and test split with  $\sim84K$, $\sim10K$, and $\sim11K$ samples, respectively. Our experiment involves the 35-class classification task. 

\begin{table}[htb!]
\centering
  \caption{Speech Commands V2 (35 classes), accuracy with model size and inference time measured on P-100 machine}
  \label{tab:speechcommands}
  \begin{tabular}{cccccc}
    \toprule
    Model & e2e & Pretrained type & Result$[\%]$ & $\#$Parameters$[\times10^6]$ & time$[msec]$\\
    \midrule
    AST \cite{gong2021ast} & \xmark & ImageNet & 98.11 & 87.3 & 11\\
    HTS-AT \cite{chen2022hts} & \xmark & AudioSet & 98.0 & 31.0 & -\\
    EAT-S & \cmark & none & \textbf{98.15} & \textbf{5.3} & 7.5\\
  \bottomrule
\end{tabular}
\end{table}

\noindent In Table \ref{tab:speechcommands}, we see that our approach achieves SotA results even without using external data, while being at least $\times 6$ lighter than other methods. Furthermore, demonstrating that our method is robust to additional content, such as speech. 

\subsection{AudioSet}
\textit{AudioSet} \cite{gemmeke2017audio} is a collection of over $2$ million $10$-second audio clips excised from YouTube videos with a class ontology of $527$ labels covering a wide range of everyday sounds, from human and animal sounds, to natural and environmental sounds, to musical and miscellaneous sounds. The set consist of two subsets, named \textit{balanced} with $20K$ samples and \textit{unbalanced} training with $2M$ samples, with evaluation set with $20K$ samples. The noise augmentations were excluded during training due to the presence of noise, and pink noise in class labels.

\begin{table}[htb]
\centering
  \caption{Audioset, mAP with model size and inference time measured on P-100, w/o external data}
  \label{tab:audest}
  \begin{tabular}{ccccccl}
    \toprule
    Model & e2e & Pretrained & mAP$[\%]$ & $\#$Parameters$[\times10^6]$ & time$[msec]$\\
    \midrule
    AST \cite{gong2021ast} & \xmark & none  & $36.6$ & $88.1$ & $63.5$\\
    ERANN-1-6 \cite{verbitskiy2021eranns} & \xmark & none  & \textbf{45.6} & 54.5 & 14\tablefootnote{Measured on V-100 machine \cite{verbitskiy2021eranns}}\\
    AemNet WM$1.0$ \cite{lopez2021efficient} & \cmark & none  & 33.16 & \textbf{5} & -\\
    HTS-AT \cite{chen2022hts} & \xmark & none & \textbf{45.3} & 31 & -\\
    EAT-S & \cmark & none  & $40.5$ & \textbf{5.3} & 8.4\\
    EAT-M & \cmark & none & 42.6 & 25.5 & 14.6\\
  \bottomrule
\end{tabular}
\end{table}

\noindent Table \ref{tab:audest}, evidently demonstrates the advantage of training a large model, for fitting large sets. Yet, our method can be a considered a good balance in terms of accuracy vs efficiency, without apparent affect on downstream tasks.
\subsection{Efficiency and edge deployment}
In this section the focus will be on model complexity. Complexity translates to model size and inference time, which can induce costs and inflexibility for platforms and applications. Our EAT-S model has 5.3M parameters, which resembles the proportion of MobileNet-V2 architecture \cite{sandler2018mobilenetv2} in both size and inference time, as detailed in Appendix \ref{app:time}. This makes EAT-S a candidate for deploying audio classification capabilities in low-memory edge devices.

\subsection{Ablation study}
In this section we explore the impact of our suggestions for augmentations and architecture. The experiments were conducted on the ESC-50 dataset.

\begin{table}[htbp]
\caption{Ablations - Classification results conducted on ESC-50 (incremental improvements over baselines)}
\label{tab:ablations}
\centering
\subfloat[Baseline $83\%$ (without any mix)]{%
    \begin{tabular}{cc}
     \toprule
    Model & Relative accuracy \\ & to baseline$[\%]$\\
    \midrule
    +mixup  & $+3.5$ \\
    +cutmix & $+0.5$\\
    +freqmix & $+3.1$\\
    +phasemix  & $+0.9$\\
    \bottomrule
\end{tabular}}%
\qquad\qquad
\subfloat[Baseline $80.5\%$ (without architecture modification)]{%
\begin{tabular}{cc}
    \toprule
    Block & Relative accuracy \\ & to baseline$[\%]$\\
    \midrule
    +modified residual blocks & $+1.8$ \\
    +dilated residual blocks & $+6.5$\\
    +transformer & $+2$\\
    \bottomrule
\end{tabular}}
\end{table}

\noindent For Tables \ref{tab:ablations}a and \ref{tab:ablations}b , in each experiment, we incrementally add to the baseline and report the relative result. 
For the augmentation ablation study, Table \ref{tab:ablations}a, the baseline refers to not applying any mixing augmentations, while in the architecture ablation study, Table \ref{tab:ablations}b, the baseline refers to our architecture without the suggested modifications - (a) modified residual block and dilated residual blocks vs. common residual block and (b) transformer vs. convolution layer with global average pooling.
We can see from Tables \ref{tab:ablations}a and \ref{tab:ablations}b that in both cases, the increments significantly improve the accuracy.

\section{Conclusions}
In this paper, we presented new audio augmentations and a novel, simple and efficient architecture for sound classification. We were able to show, through analysis and experiments, that end-to-end audio systems can no longer be considered inferior, especially in low data scenarios. The suggested scheme achieves state of the art results in several datasets both in 'from-scratch' and AudioSet-pretraining setups, all while being exceptionally light-weight and robust. Future work can elaborate this work to solve additional tasks and contents, such as sound event detection, localization or speech and speaker recognition.
\bibliographystyle{unsrt}  
\bibliography{references}  
\appendix
\section*{Appendices}
\section{Phase waveform synthesis}\label{app:phase}
We assumed that the audio signal's phase component may contain discriminative information. To confirm this assumption, we synthesized waveforms from the phase of the signal. Spectrogram based representations inherently ignore the phase, by taking only the magnitude. Mathematically this is equivalent to filtering the original signal with a unit amplitude filter with the opposite phase, as described at equations $[$\ref{eq:phase1}, \ref{eq:phase2}$]$:

\begin{align}
    \label{eq:phase1}
    X[k,l] &= STFT\left(x[n]\right)\nonumber\\
    X[k,l] &= \left|X[k,l]\right|\cdot e^{j\phi[k,l]}\nonumber\\
    \left|X[k,l]\right|&=X[k,l] \cdot e^{-j\phi[k,l]}
\end{align}
Which can rephrased as filtering operation - 
\begin{align}
    \label{eq:phase2}
    \left|X[k,l]\right| &= X[k,l] \cdot H[k,l]\nonumber\\
    h[n] &= ISTFT(H[k,l])
\end{align}
The "phase" waveform was extracted according to equation \ref{eq:phase2}. For this experiment, the classifier was trained (a) on the original signal, (b) signals based on phase, (c) signals based on the magnitude, and (d) on both (b)+(c), concatenated to produce a 2-dimensional input signal. The experiment was conducted on the ESC-50 dataset, with noise and mixing augmentations being disabled \footnote{Due to the noisy nature of the phase signal}.

\begin{table}[htb]
  \centering
  \caption{ESC-50, accuracy vs input content}
  \label{tab:phonly}
  \begin{tabular}{ccl}
    \toprule
    Mode & Accuracy$[$\%$]$\\
    \midrule
    phase & 60\\
    magnitude & 78.5\\
    phase+magnitude & 80.5\\
    baseline & 81\\
    \bottomrule
\end{tabular}
\end{table}

From Table \ref{tab:phonly} we can see two outcomes. At first, the phase signal resulted in significantly higher accuracy than a random guess. Second, seems that this information is complementary to the magnitude signal. These observations lead us to augment the phase domain during the training process, by adding phase noise and mixing the phases among pairs of signals.

\section{Inference time details}\label{app:time}
\setcounter{section}{2}
Table \ref{tab:inference} details the inference time for various network configurations. 

\begin{table}[htb]
\centering
  \caption{Inference time measured on V-100 machine, and Intel(R) Xeon(R) CPU E5-2682 v4 @ 2.50GHz on EAT-S model}
  \label{tab:inference}
  \begin{tabular}{ccc}
    \toprule
    sample length$[s]$ & gpu-time$[ms]$ & cpu-time$[ms]$\\
    \midrule
    1 & 5.3 & 38.3\\
    5 & 5.5 & 67\\
    10 & 5.6 & 145\\
    \bottomrule
\end{tabular}
\end{table}

\section{EAT Models - Architectures details}\label{app:arch}
\setcounter{section}{3}
Table \ref{tab:EATsize} details the configuration for our models, EAT-S/M. "Channels" refers to number of filter at the first stage of the network.
\begin{table}[H]
  \centering
  \caption{Architecture details}
  \label{tab:EATsize}
  \begin{tabular}{ccccc}
    \toprule
    Model & Channels & Transformer layers/heads & embedding dimension & $\#$Parameters$[\times10^6]$\\
    \midrule
    EAT-S(mall) & 16 & 4/8 & 128 & 5.3\\
    EAT-M(edium) & 32 & 6/16 & 256 & 25.5\\
  \bottomrule
\end{tabular}
\end{table}
\end{document}